\documentstyle[aps]{revtex}
\input epsf
\begin{document} 

\newcommand{\be}{\begin{equation}}
\newcommand{\ee}{\end{equation}}

\draft

\title{Preparation of
decoherence-free, subradiant states in a cavity}

\date{\today}

\author{
P\'eter F\"oldi\cite{PFemail}, 
Mih\'{a}ly G. Benedict\cite{MGBemail}
Attila Czirj\'{a}k\cite{ACemail}, 
}

\address{
Department of Theoretical Physics, University of Szeged, 
H-6720 Szeged, Tisza Lajos krt. 84-86, Hungary
}
\maketitle

\begin{abstract}
The cause of decoherence
in a quantum system can be traced back to the interaction with the 
environment. As it has been pointed out first by Dicke, in a system of $N$
two-level atoms
where each of the atoms is individually dipole coupled to the environment,   
there are collective, subradiant states, that have no dipole 
coupling to photon modes, and therefore they are expected 
to decay slower.  This  property also implies that these type 
of states, which form an $N-1$ dimensional subspace of the atomic subsytem, 
also decohere slower. 
We propose a scheme which will create such states. 
First the two-level atoms are placed in a strongly 
detuned cavity and one of the atoms, called the control atom is excited.
The time evolution of the coupled   
atom-cavity system leads to an appropriately entangled state of the atoms. 
By applying subsequent laser pulses at a well defined time instant,
it is possible to drive the atomic state into the subradiant, i. e., 
decoherence free subspace. Up to a certain average number of the photons, 
the result is independent of the state of the cavity.
The analysis of the conditions shows that this scheme is 
feasible with present day techniques achieved in atom cavity 
interaction experiments.
\end{abstract}


%%%%%%%%%%%%%%%%%%%%%%%%%%%%%%%%%%%%%%%%%%%%%%%%%%%%%%%%%%%%%%%%%%%%%%%%%%%%%%%%%%%%%%%%%%%%%%%%

Subradiant states of a system of two-level atoms 
\cite{dicke,SRAD,grhar982,crub985,devoe996} has
recently gained wide attention because of their exceptionally
slow decoherence \cite{decref1,decref2,fcb001}.  
This stability of quantum superpositions inside the subradiant
subspaces originates from the low probability of photon emission, 
which means very weak interaction between the
atoms and their environment.    
Hence the subradiant states span a
decoherence-free subspace (DFS) \cite{lidar998,knight999,knight000}
of the atomic Hilbert-space and
consequently can become important from the viewpoint of quantum computation
(QC)\cite{knight999,divin000}.
The scheme we propose can be used to prepare subradiant states 
in a cavity. Our method is based on second order perturbation
theory but the exact results verify the
validity of the perturbative approach.
We also investigate to what extent our scheme
is independent of the state of the cavity field.
Finally the requirements needed to prepare subradiant
states in the proposed way will be compared with 
available experimental techniques \cite{walt999,brhar996,har997,har999}.

We investigate a system of $N$ identical two-level atoms in a single mode
cavity. Each individual atom is equivalent to a spin-$1/2$ system, and
the whole atomic ensemble can be described by the aid of 
collective atomic operators  $J_+$, $J_-$ and $J_z$
obeying the same algebra as the usual angular momentum 
operators\cite{dicke}.
We consider the following model Hamiltonian:
\begin{equation}
H=H_0+H_{int}=\hbar \omega_{\mathrm{a}}J_{z}+ \hbar \omega _{c}a^{\dagger
}a+ \hbar g \left( a^{\dagger }J_{-}+aJ_{+} \right),
\label{model_ham}
\end{equation}
where $a$ and $a^\dagger$ are the annihilation and creation operators 
of the cavity mode, $\omega_{\mathrm{a}}$ is the transition frequency 
between the two atomic 
energy levels,  $\omega_c$ denotes the frequency of the cavity 
mode, different from 
$\omega_{\mathrm{a}}$, and $g$ is the 
coupling constant. 
We note that the Hamiltonian (\ref{model_ham}) 
is written in the framework of Dicke's theory, i.e.,
with the assumption 
that all the atoms are subjected to the same field, which is a good 
approximation when the size of the atomic sample is small compared to
the wavelength of the cavity mode. As discussed later in detail, 
there are experimental situations where this
requirement is fulfilled.
Our proposed scheme for preparing subradiant states
involves a detuned cavity. We shall assume that the detuning is much larger
than the resonant Rabi frequency:
\begin{equation}
\omega_c-\omega_a= \Delta \gg g.
\end{equation}

Now any state  of the atomic system and the cavity field
can be expanded as a linear combination of eigenstates of $H_0$.  
These are tensorial products
of collective atomic states and number states of the field: 
$|j,m,\lambda\rangle \otimes |n\rangle$, where the indices 
$j,m$ and $\lambda$ label the 
atomic state (also called Dicke state \cite{acgt972}) 
while $n$ refers to the {\it n}th Fock state of the
mode. 
The quantum number $j$ corresponds to
 the eigenvalues of the
operator $J^2=J_z^2+(J_+J_-+J_-J_+)/2$. This index is in one-to-one
correspondence with the Young digram
\cite {hamermesh} that describes the permutation 
symmetry of the state. The possible values of $j$ is $N/2, N/2-1,\dots $,
the smallest value being  $0$ if $N$ is even and $1/2$ if $N$ is odd.
The index $m$ of the $|j,m,\lambda\rangle$
Dicke state labels the eigenstates of the collective atomic operator 
$J_z$, that is essentially proportional to the 
energy of the atomic subsytem.
This is the index that is decreased (increased) by one under 
the action of the operator $J_-$ ($J_+$):
\be
J_-|j,m,\lambda\rangle=\sqrt{j(j+1)-m(m-1)}|j,m-1,\lambda\rangle,
\label{jpaction}
\ee
including the case when $m=-j$, when the result is the zero vector.
The states with $m=-j$ are the lowest ones of the Dicke ladders
\cite{dicke}, they are called subradiant, because they have 
no diploe coupling to other lower lying states.    
Finally the index $\lambda$ distinguishes states with the
same $j$ and $m$.  
For more details see 
Refs. \cite{dicke,fcb001,hamermesh,acgt972,crub987,keitscully992,bc999}.

Besides the collective atomic states $|j,m,\lambda\rangle$, we shall 
also use the natural basis that assigns a well defined state to each  
individual atom. These vectors will be labeled by a string of $0$-s and $1$-s
corresponding to the ground and excited sates, respectively.
E. g., the ground state of the atomic subsystem
is written in this basis as $|\stackrel{1}{0} \stackrel{2}{0}
\ldots \stackrel{N}{0}\rangle$; this state (as well as the fully
excited one) is also 
an element of the Dicke basis, $|00\ldots 0\rangle=
|j=N/2,m=-N/2,\lambda=1\rangle$.

The form of the Hamiltonian (\ref{model_ham}) implies that
the time evolution of the system shall exhibit two time 
scales: The first characteristic time is due to the self-Hamiltonian $H_0$ and
is approximately $2\pi/\omega_a$ (or $2\pi/\omega_c$) and
the second is proportional to $2\pi/g$. Generally $g \ll 
\omega_a \approx \omega_c $ 
and the faster process induced by $H_0$ can be eliminated by going into
an interaction picture. However, if the frequency difference $\Delta$ is large
enough, then the energy transfer between two adjacent eigenstates of $H_0$, 
differing in only one photon number, becomes negligible. 
This means that the amplitude of
the corresponding collective Rabi oscillations will be very small, that is,
the  process 
on the second time scale will be unnoticeable and even slower mechanisms will 
become apparent. The situation is similar to the proposals \cite{sormol999}
and \cite{zhguo000}.

Hereafter we shall focus on the solution of 
the Schr\"{o}dinger equation
in the case when just a single atom is excited at $t=0$.
This initial state can be prepared by  
starting from the state $|00\ldots0\rangle$,
and exciting one well defined control atom.
This excitation can be achieved via a third much higher lying level,
so that the wavelength of the addressing pulse allows to focus it on the
desired target atom \cite{blatt999}.
For the sake of simplicity we always consider the control atom as being the 
first, hence the initial state will be written as 
\be
|\phi(0)\rangle=|100\ldots 0 \rangle \otimes|n-1\rangle.
\label{veryfirst}
\ee 

In order to find the complete analytical solution of the 
Schr\"{o}dinger equation
induced by the Hamiltonian (\ref{model_ham}), in principle one should 
calculate all the eigenvalues and the corresponding eigenstates of $H$. 
Athough this problem can be solved analytically \cite{TC68}, 
more insight is given by a simple perturbative
approach. 
The exact nonperturbative numerical solution of the Schr\"{o}dinger equation
verifies that results obtained via perturbation theory yield  excellent  
approximations.  

The state 
\begin{eqnarray} 
|1\rangle \equiv \left({{1}\over{\sqrt{N}}}
\sum_{k=1}^N |0\ldots 0\stackrel{k}{1}0 
\ldots 0\rangle  \right)\otimes|n-1\rangle= \nonumber \\
= |j=N/2,m=-N/2+1,\lambda=1\rangle\otimes|n-1\rangle ,
\label{symmetric}
\end{eqnarray}
which is in the completely symmetric subspace, and the 
subradiant states: 
\begin{equation} 
|i\rangle \equiv |j=N/2-1,m=-N/2+1,\lambda=i-1\rangle \otimes|n-1\rangle
\label{kn}
\end{equation}
 with $i=2,3...N$,  
have the same unperturbed energy, they span the $N$-fold degenerate 
eigensubspace 
of $H_0$ corresponding to the eigenvalue 
$E^0(n)=\hbar(n \omega_c-N\omega_a/2)-\hbar\Delta$. 

It can be seen that first order degenerate perturbation theory is 
not giving any correction to the energy, because all the matrix elements of
$H_{int}$ between the states above vanish, the action 
of $H_{int}$ on vectors $|j,m,\lambda\rangle\otimes|n-1\rangle$ 
gives a linear combination of $|j,m-1,\lambda\rangle\otimes|n\rangle$ and 
$|j,m+1,\lambda\rangle\otimes|n-2\rangle$ that are orthogonal
to the states (\ref{symmetric}) and (\ref{kn}).
In order to obtain nonzero energy corrections we have to make a second order 
degenerate perturbation calculation\cite{landau},  and find the eigenvalues 
of the matrix: 
\be
\sum_m {{\langle i|H_{int}|m\rangle\langle m|H_{int}|k\rangle}
\over{E^0(n)-E_m^0}},
\label{secular2}
\ee
where the sum runs over all eigenstates of $H_0$ with eigenvale
$E_m^0\neq E^0(n)$.
The only nonvanishing energy corrections in second order are the following:
\begin{eqnarray}
\delta E_1&=&\hbar {{g^2}\over{\Delta}}(Nn-2N-2n+2),\nonumber \\ 
\delta E_i&=&\delta E_1+\hbar N{{g^2}\over{\Delta}}, \ \  i=2,3\ldots N.
\label{corrections}
\end{eqnarray}

At this point we can formulate the requirements that assure the
validity of the perturbation theory: the magnitude of 
$\delta E_1$ and $\delta E_i$ must be much smaller than 
$\hbar |\Delta|$, the minimum of the
difference between $E^0(n)$ and all other unperturbed energy levels. 

The most important consequence of Eqs. (\ref{corrections}) 
is that the Bohr frequencies that determine the 
time dependences
of the subradiant and non-subradiant states are different.
 
Now we expand the initial state (\ref{veryfirst}) 
as the linear combination of the fully 
symmetric (non-subradiant)
state $|1\rangle$,
and an appropriate subradiant state:
\begin{eqnarray}
|2\rangle={{1}\over{\sqrt{N(N-1)}}}\Big[(N-1)|100\ldots 0 \rangle- 
\nonumber \\ 
\sum_{k=2}^N |0\ldots 0\stackrel{k}{1}0 
\ldots 0\rangle  \Big]
\otimes|n-1\rangle.
\label{subradiant}
\end{eqnarray}
By assigning the symbol $|2\rangle$ 
to the state in Eq. (\ref{subradiant}), we have utilized the freedom
of choosing a basis in the subradiant subspace. 
Now the initial state reads 
\be
|\phi(0)\rangle={{1}\over{\sqrt{N}}}|1\rangle+\sqrt{{N-1}\over{N}}|2\rangle.
\label{expansion}
\ee 
By the aid of this expansion and using the Bohr frequencies 
resulting from (\ref{corrections}), it is easy to calculate the 
time evolution of the state (\ref{expansion}).
Discarding an overall phase factor, this time dependent state has the form
\be
|\phi(t)\rangle={{1}\over{\sqrt{N}}}\exp\left(i N {{g^2}\over{\Delta}} 
t \right)|1\rangle
+\sqrt{{N-1}\over{N}}|2\rangle,
\label{expevol}
\ee
or, on using Eqs. (\ref{symmetric}) and  (\ref {subradiant}):
\begin{eqnarray}
|\phi(t)\rangle= 
\Big[ \left( N \cos(\alpha t)-i(N-2)\sin(\alpha t)\right)|100\ldots 0 \rangle
\nonumber \\
+2 i\sin(\alpha t) \sum_{k=2} |0\ldots 010 \ldots 0\rangle \Big] 
\otimes|n-1\rangle /N.
\label{updownevol}
\end {eqnarray}
Here we introduced the parameter
\be 
\alpha={{N g^2}\over{2\Delta}},
\label{gamma}
\ee 
which is {\it independent} of $n$.
Because of this latter fact, from now on 
the state of the cavity field will be omitted in the notation.
We also note that the characteristic time 
of the time evolution, ${{2\pi}/{\alpha}}$, is much longer
than that of the free evolution due to $H_0$, being the consequence 
of the fact that the evolution described in Eq. 
(\ref{expevol}) is induced by a weak, nonresonant
interaction.

Eq. (\ref {updownevol})  reveals that in $|\phi(t)\rangle$ the weight of the 
state  $|100\ldots 0 \rangle$ and those of the states with the first atom unexcited 
changes during the course of time. As we can see, the moduli 
of the corresponding coefficients in Eq. (\ref
{updownevol}) are 
$$
 {{\sqrt{N^2 \cos^2(\alpha t)+(N-2)^2\sin^2(\alpha t)}}\over{N}}
\ \ 
{\rm and}
\ \  
{{2|\sin(\alpha t)|}\over{N}},
$$
respectively. Comparing these values to Eq (\ref{subradiant}), it can be 
shown that for an arbitrary  $N$ there exists a time instant $t_m$ 
when
\begin{eqnarray}
|\phi(t)\rangle={{1}\over{\sqrt{N(N-1)}}}\Big[(N-1)e^{i \varphi}
|100\ldots 0 \rangle- 
\nonumber \\ 
\sum_{k=2}^N |0\ldots 0\stackrel{k}{1}0 
\ldots 0\rangle  \Big],
\label{correctmod}
\end{eqnarray}
which differs from the subradiant state $|2\rangle$
only in the phase factor $e^{i\varphi}$ of the first term.
Combination of the previous two equations and Eq. (\ref{subradiant})
yields the following requirement for $t_m$:
\be
{{\sqrt{N^2 \cos^2(\alpha t_m)+(N-2)^2\sin^2(\alpha t_m)}}
\over{\left| 2\sin(\alpha t_m)\right|}}=N-1.
\ee 
We can find a solution of this equation for all $N>1$:
\be
\sin \alpha t_m = \sqrt{N/(4N-4)},
\label{tmod}
\ee
and also obtain 
$\cos \varphi={{N-2}\over{2N-2}}$ in Eq. (\ref{correctmod}).

Now it is clear that at the time instant given by Eq (\ref{tmod}), 
an appropriate 
rapid change in the phase of the state $|100\ldots 0 \rangle$ 
relatively to all other states $|0\ldots 0\stackrel{k}{1}0 
\ldots 0\rangle$  leads to the subradiant state $|2\rangle$.

On the other hand, Eq. (\ref{correctmod}) also shows that
the required phase transformation is equivalent to the
elimination of the phase difference $\varphi$ between
the $|1\rangle_c$ excited and $|0\rangle_c$ ground state of 
the control atom.
Therefore we consider the action of a strong laser pulse on the
control atom. In order to obtain precise addressing \cite{blatt999}, the 
laser is to be tuned in resonance with an allowed transition
$|1\rangle_c \rightarrow |e\rangle_c$, where $|e\rangle_c$
denotes a state of the control atom with much higher energy
than $|1\rangle_c$. E. g., by the appropriate choice of the phase
of the complex Rabi frequencies of two $\pi$ pulses leads to
the phase transformation reqired to prepare the subradiant 
state $|2\rangle$.   
Additionally, the duration of a Rabi period 
due to the strong, resonant laser pulse
is much shorter than the characteristic time
that governs the time evolution (\ref {expevol}).
We note that the idea of introducing phase transformation 
in a multilevel system by the aid of
short laser pulses has appeared in a somewhat different context
in \cite{ASW001}. 

Now we show that our scheme is independent of the state of the 
cavity field, and write more generally the initial state as: 
\be
|\phi(0)\rangle=|100\ldots 0 \rangle \otimes|\psi(t)\rangle=
|100\ldots 0 \rangle \otimes \sum_n c_n(t) |n\rangle.
\label{nonfock}
\ee
We use the fact that
the interaction Hamiltonian $H_{int}$ does not mix states 
with different number of excitation (essentially $n+m$): 
\be
\langle j,m,\lambda| \otimes \langle n| H_{int} |n^{\prime}\rangle
\otimes |j,m\pm 1,
\lambda\rangle =0,
\label{nonmix}
\ee
unless $n^{\prime}=n \mp 1$.  
This implies  that the calculations based on second order
perturbation theory can be performed for each $N$-fold degenerate 
energy level of $H_0$ corresponding to different values of $n$. 
After replacing the state $|n-1\rangle$ 
with $|\psi (t)\rangle$ in Eqs. (\ref {symmetric}) and (\ref {kn}), 
we obtain the following result:
\be
\delta E_i-\delta E_1
={{g^2}\over{\Delta}} N \sum_n |c_n|^2=2\alpha,
\ee
which is therefore also valid in this general case.
Thus we have proven that our scheme does not require
special preparation of the cavity field.
However, it should be borne in mind, that the results above are based on 
perturbation theory. For given $N, g$ and  $\Delta$ the
validity of the perturbative calculations depends on 
$\langle n \rangle$, the average number of photons in the cavity field.   
Hence it is clear that our scheme can not be independent of the 
average photon number on a very large scale. 
Nevertheless, until ${{g}\over{\Delta}}
\sqrt{N\langle n \rangle} \ll 1$, all the previous statements hold.
For the case of $10$ atoms, we have performed exact 
(nonperturbative) numerical calculations
for the experimentally realizable  ratio \cite{brhar996} of $g/\Delta=30$,
and found that the time evolution follows
Eq. (\ref{expevol}) within $2\%$ relative error in the
coefficients.  

Finally we compare the requirements of our scheme
with the experimental possibilities.    
The atom cavity experiments of Haroche and co-workers, 
as described in the review paper \cite{HR85}, show that the description of 
the interaction of a number of Rydberg atoms with a single mode cavity 
is truly described in the framework of the Dicke model.
In more recent experiments 
\cite{brhar996,har997,har999} the interaction of a 
detuned cavity with one and two atoms has been found to be in 
agreement with theoretical predictions.
The parameters realized in these   
experiments with rubidium Rydberg atoms: $g/2\pi\approx 24 kHz$ 
and detunings 
as large as $\Delta/2\pi \approx 800 kHz$ show that the conditions of the
validity of our perturbation approach hold as much as for about 
hundred atoms, because the average photon number in a cavity can be 
kept much less than 1. 
Taking for instance $10$ atoms and 
$g/\Delta=30$, we have $\alpha={2.5 \times 10^4} 1/s$ giving    
for $t_m$ a value of $22 \mu s$. 
The interaction time of the atoms and the cavity
must be longer than $t_m$ what can be achieved already 
with atoms with somewhat less than thermal velocities for 
centimeter sized cavities.  
Finally we note that addressing of single atoms is a 
common problem in almost all 
of the proposals in QC, but there are 
promising works indicating future success \cite{blatt999}. 

In conclusion, we have proposed a method to prepare 
decoherence-free, subradiant
states of a multiatomic system.  
We have shown that our perturbative
approach is compatible with present day techniques in atom cavity
experiments. 

We thank S. B. Zheng for discussions. 
This work was supported by the Hungarian Scientific Research
Fund (OTKA) under contract T32920, and by the Hungarian Ministry of
Education under contract FKFP 099/2001.

\end {document}